\documentclass[aps,prd,prabib,twocolumn,showpacs,nofootinbib]{revtex4}
\usepackage{graphicx} \usepackage{amsmath} \usepackage{amssymb}
\usepackage{amsfonts} \usepackage{bm}

\begin{document}

\newcommand{\be}{\begin{equation}} \newcommand{\ee}{\end{equation}}
\newcommand{\bea}{\begin{eqnarray}}\newcommand{\eea}{\end{eqnarray}}

\title{A Family of Non-commutative  geometries}

\author{Debabrata Sinha} \email{debabratas@bose.res.in}

\affiliation{Satyendra Nath Bose  National Centre for Basic Sciences,
Block-JD, Sector-III, Salt Lake, Kolkata - 700098, India}

\author{Pulak Ranjan Giri} \email{pulakgiri@gmail.com}

\affiliation{Physics and Applied Mathematics Unit,
Indian Statistical Institute, Kolkata 700108, India}

\begin{abstract}
It is shown that  the non-commutativity in quantum Hall system may get
modified. The self-adjoint extension
of the corresponding Hamiltonian leads to a family of non-commutative
geometry labeled by the self-adjoint extension parameters. We explicitly
perform an exact calculation using a singular interaction and show that,
when projected to a certain Landau level,
the  emergent non-commutative geometries of the projected coordinates belong to a one parameter family. There is a possibility of obtaining  the filling fraction of fractional quantum Hall effect  by suitably choosing  the value of the self-adjoint extension parameter.
\end{abstract}

\pacs{03.65.-w, 02.40.Gh, 73.43.-f}


\date{\today}

\maketitle


Non-commutative geometry has been one of the contemporary issues in recent
research fields \cite{michael,calmet,pulak1,pulak2} in theoretical physics. The simplest non-commutativity one assumes  is of the form
\begin{eqnarray}
[x_\mu,x_\nu]= i\theta_{\mu\nu}\,,
\label{non1}
\end{eqnarray}
where the real constant $\theta_{\mu\nu}$ is the strength of non-commutativity.
It is believed that the
effect of spacetime non-commutativity would be important in Plank length scale, i.e.,
$\theta_{\mu\nu}\sim l_P^2$, which is far  from present day experiments. However one
may look into the low energy sector to find some form of non-commutativity.
The realization of non-commutative spatial geometry in quantum Hall effect \cite{my,big,BC,sus,nair,hor1} is one of them. The key ingredient in  quantum Hall effect is the  presence of landau levels. One can show
that the $2$-dimensional spatial geometry in the Lowest landau level becomes noncommutative.
The strength of non-commutativity is in general inversely proportional to the magnetic field
B applied in the system, implying
\begin{eqnarray}
[x,y]= i\frac{1}{B}\,.
\label{non2}
\end{eqnarray}
One can constrain the system in lowest landau level by taking the zero mass limit \cite{dunne1,dunne2}, $m\to 0$,
of the evolving particle or by making the magnetic field very large \cite{vei,my1}, $B\to \infty$.
In these limit the coordinates of a plane becomes conjugate to each other.
In this view interaction  and/or boundary conditions does not seem to have any effect
on the non-commutative structure. However there is  another way to achieve non-commutativity
in the Landau system.  Instead of taking limit on the mass or the applied magnetic field one can
project the coordinates on the lowest or any specific  Landau level. Then the projected coordinates  $x_P, y_P$  becomes conjugate to each other \cite{mac} and in some units satisfy
\begin{eqnarray}
[x_P,y_P]= i\frac{1}{B}\,,
\label{non3}
\end{eqnarray}
This point of view is important for the study of  fractional quantum Hall effect.
Because, if we take into account the interaction  between the electrons then the non-commutative
structure will in general change. Suppose the interaction between the electrons are of the form
$V_I= \lambda^2\mathcal{F}(x)$ and suppose the Hamiltonian of the Landau problem remains
essentially self-adjoint with the introduction of the interaction. Then the  commutator
of the projected coordinates will get modified as
\begin{eqnarray}
[x_P,y_P]= i\Theta(B,\lambda)\,,
\label{non4}
\end{eqnarray}
where the real number $\Theta(B,\lambda)$  is now the strength of non-commutativity which depends on the coupling constant  $\lambda$ of the interaction and the magnetic field $B$ in a complicated way. The exact form of the non-commutativity have to be determined by explicitly solving the corresponding Schr\"odinger equation. Of course in the limit of vanishing interaction  one may obtain
\begin{eqnarray}
\lim_{\lambda\to 0}\Theta(B,\lambda)= \frac{1}{B}\,,
\end{eqnarray}
which gives the familiar result  (\ref{non3}). The effect of interaction on the non-commutativity may have importance in understanding the fractional quantum Hall effect. This can be understood as described in  \cite{BC} by calculating the minimal area acquired by a particle in the non-commutative plane.
Note that the  non-commutativity (\ref{non3}) implies that the minimal area in the projected space to be
\begin{eqnarray}
\Delta A\equiv\Delta x_P \Delta y_P\simeq \frac{1}{B}\,,
\label{non5}
\end{eqnarray}
Then for an area  $A$ number of states available to the electrons in a Landau level is
$M= \frac{A}{\Delta A}= AB$. The filling  fraction for a system of $N$ number of electrons is given by
\begin{eqnarray}
\nu= \frac{N}{M}= \frac{1}{B}\frac{N}{A}\,,
\label{fill1}
\end{eqnarray}
For integral quantum Hall effect  $\nu$ is some integer  and can be described by noninteracting electrons.
However  for fractional quantum Hall effect interactions between the electrons are important.
One should therefore consider the modified non-commutativity described by  (\ref{non4}),
which then implies  the filling fraction to be
\begin{eqnarray}
\nu= \frac{N}{M}= \Theta(B,\lambda)\frac{N}{A}\,,
\label{fill2}
\end{eqnarray}
Note that the above conclusion is based on  the assumption that the Hamiltonian is essentially self-adjoint.
However in actual physical situation the interaction between the electrons may  make the Hamiltonian
non-selfadjoint  at least for the s-waves which is important for fractional quantum Hall effect.
We therefore in this article assumes  that the addition of interaction makes the the Hamiltonian
\begin{eqnarray}
H_L= (\bold{p} +e\bold{A})^2/2m_e \,,
\label{ha1}
\end{eqnarray}
of the Landau problem non-selfadjoint but it has one parameter family of self-adjoint extensions.
Here $\bold{A}$ is the magnetic vector potential corresponding to the constant
magnetic field $B$ perpendicular to the plane and $m_e$ is the reduced mass of two electron system.  The interaction potential between the electrons we consider
is of the form
\begin{eqnarray}
V_I= \frac{\lambda^2}{r^2}\,,
\label{pot1}
\end{eqnarray}
This interaction has a similarity with the gauge potential $A_i= -\frac{\lambda^2}{r^2}\epsilon_{ij}x^j$,
which corresponds to a singular flux tube situated at the origin of the coordinates.
This kind of singular potential is important to explain quantum Hall effects.
The eigenvalue equation
$H\psi=E\psi$  will  govern the shifts of the Landau levels due to the  interact  potential $V_I$, where
\begin{eqnarray}
H= H_L + V_I\,.
\label{ham1}
\end{eqnarray}
The introduction of the potential $V_I$ changes
the  short distance behavior of the wave-functions $\psi$,
which  is responsible for making the Hamiltonian non-self-adjoint. This can be understood from Weyl's limit point-limit circle (LPLC) theory.

Before we actually study the Landau problem with interaction, we here give a brief discussion of the LPLC method. Elaborate discussion on it  can be found in the book of Reed and Simon \cite{reed}. For an ordinary  second order linear differential equation of the form
\begin{eqnarray}
H_W\psi_W\equiv \left(-\frac{d^2}{dx^2} + V_W\right)\psi_W=
E_W\psi_W\,,\label{limit1}
\end{eqnarray}
defined in $C^\infty_0(0,\infty)$, the potential $V_W$  has the following characteristics. It is said to be in the limit circle case at
zero or at infinity respectively, if for all $E_W$ all solutions of
(\ref{limit1}) are square-integrable at zero or at infinity
respectively. If $V_W$ is not in the limit circle
case at any of the two boundaries then it is in limit point case at
that boundary.  The Hamiltonian $H_W$ is essentially self-adjoint
on $C^\infty_0(0,\infty)$ if and only if $V_W$ is in the limit
point case at both ends, zero and infinity.
It the potential $V_W$ is in the
limit circle case at both ends then the
deficiency indices of the Hamiltonian $H_W$ are both same,
$n_+=n_-=2$.  Note that the deficiency indices $n_\pm$ are the  number of
solutions $\psi^\pm_W$ of the deficiency equations
\begin{eqnarray}
\left(-\frac{d^2}{dx^2} + V_W\right)\psi^\pm_W=
\pm \beta\psi^\pm_W\,,
\end{eqnarray}
where $\beta$ is any complex numbers, but for calculation purpose we will take $\beta=i$. The
Hamiltonian, for which $n_+=n_-=2$, is not self-adjoint but admits $4$-parameter family of
self-adjoint extensions. Another situation is when $V_W$ is in the
limit point case at one boundary point but limit circle case at
another boundary point, then the deficiency space solutions are both
same. But this time they are all one, i.e., $n_+=n_-=1$. This time
however one can have a one parameter family of self-adjoint
extensions. Finally when $V_W$ is in limit point case at both ends, then the
deficiency space solutions are all zero, i.e., $n_+=n_-=0$ and  the Hamiltonian is essentially self-adjoint.

Let us now return to our  problem given by the Hamiltonian $H$. In order to apply the above method we identify the explicit form of the  potential $V_W$ to be
\begin{eqnarray}
V_W= \frac{\sigma^2 -\frac{1}{4}}{r^{2}}+\frac{1}{4}\omega_B^{2}r^{2}\,,
\end{eqnarray}
in radial coordinates, where $\sigma^2= l^2+\lambda^2$ and $\omega_B=B/2$. The corresponding  radial wave-function $\psi_W$ is part of the full
wave-function $\psi=\frac{1}{\sqrt{r}}\psi_W\exp(il\phi)$.

Note that the short distance behavior is dominated by the inverse square potential while the long distance behavior is dominated by the harmonic potential. The solutions at short distance is of the form
\begin{eqnarray}
\lim_{r\to 0}\psi_W \simeq  r^{(1/2\pm \sigma)}\,.
\end{eqnarray}
Both the above solutions are square integrable at $r \to 0$ if $\sigma$ lies in the interval  $\sigma \in (-1,+1)$, which then makes the potential $V_W$ in limit circle case at zero. Outside the interval the potential is in limit point case. The long distance $r\to \infty$ behavior is however unperturbed by the interacting potential $V_I$. One of the solutions is
square-integrable  and  behaves as
\begin{eqnarray}
\lim_{r\to \infty}\psi_W \simeq  e^{-\frac{1}{4}\omega_B r^2}\,.
\end{eqnarray}
The other solution is not square-integrable, which is therefore not acceptable. The potential $V_W$ is therefore  in the limit point case at $r\to \infty$. Outside the critical interval $\sigma \in (-1,+1)$ the potential is in the limit point case at both ends and therefore the the Hamiltonian is essentially self-adjoint.  However in the critical interval, since one  end is in limit circle case and other end is in limit point case, $H$ is not self-adjoint but  has a one parameter family of  self-adjoint extensions. Note that the deficiency space solutions in this case are
\begin{eqnarray}
\psi_W^\pm= r^{(\frac{1}{2}+ \sigma)}e^{-\frac{\omega_B}{4}r^2}U(\xi^\pm,1+ \sigma,\frac{\omega_B}{2}r^2)\,,
\end{eqnarray}
where $\xi^\pm =\mp \frac{i}{2\omega_B}+\frac{\sigma+1}{2}$  and $U$ is confluent hypergeometric function \cite{abr}.  Existence of one square integrable solution of both kind gives the same conclusion that Hamiltonian  has a one parameter family of self adjoint extensions when the coupling $\sigma$ is in the  critical interval. The method of finding  deficiency space solutions to construct self-adjoint extension is known as von Neumann method \cite{reed}. Some of the problems specifically inverse square problem, which is relevant in this case,  have been performed in \cite{giri,giri1}.
Given the domain $D_W$ of the symmetric operator $H_W$ the self adjoint extensions,
characterized by $ e^{i\alpha}, \alpha \in [0,2\pi]$, is  represented by the domain
\begin{eqnarray}
D_\alpha \equiv D_W +\psi_W^+ + e^{i\alpha}
\psi_W^-\,. \label{d6}
\end{eqnarray}
The radial solution $\psi_W$ explicitly can be written as
\begin{eqnarray}
\psi_W=C_l r^{(\frac{1}{2}+ \sigma)}e^{-\frac{\omega_B}{4}r^2}U(\xi,1+ \sigma,\frac{\omega_B}{2}r^2)\,,
\end{eqnarray}
where 
\begin{eqnarray}
C_l=\sqrt{\sqrt{2\omega_B}\sin(\frac{E+\omega_Bl}{2\omega_B})
\frac{\Gamma(\xi)\Gamma(1-\xi)}{\Psi(\xi)-\Psi(1-\xi)}}\,,
\end{eqnarray}
is the normalization constant explicitly depends on the eigenvalue $E$ and $\xi =-\frac{E+\omega_Bl}{2\omega_B}+\frac{\sigma+1}{2} $. In order to find out the explicit form of the eigenvalue we need to match the behavior of the solution  $\psi_W$ at $r\to 0$
\begin{eqnarray}
\lim_{r \to 0}\psi_W =A r^{(\frac{1}{2}+\sigma)}+B r^{(\frac{1}{2}-\sigma)}\,,
\label{psiw1}
\end{eqnarray}
where
\begin{eqnarray}
\nonumber A&=&\frac{\pi}{\sin\pi(1+\sigma)}\frac{1}{\Gamma(\xi-\sigma)\Gamma(1+\sigma)}\\
\nonumber B&=&\frac{\pi}{\sin\pi(1+\sigma)}\frac{1}{\Gamma(\xi)\Gamma(1-\sigma)}
\end{eqnarray}
with  $D_\alpha$.
The behavior of any function, belonging to the domain $D_\alpha$, near singularity $r\rightarrow 0$ can be found from the behavior of $\psi_W^+ + e^{i\alpha}\phi_W^-$ at short distance, because near singularity the function belonging to the domain $D_W$ goes to zero. So,we can write
\begin{eqnarray}
\lim_{r \to 0} D_\alpha=\lim_{r \to 0}(\psi_W^+ + e^{i\alpha}\psi_W^-)
\end{eqnarray}

\begin{eqnarray}
\nonumber \lim_{r\to 0}\psi_W^{+} &=& Mr^{(\frac{1}{2}+ \sigma)}+Nr^{(\frac{1}{2}-\sigma)}\\
\lim_{r\to 0}\psi_W^{-} &=&M^*r^{(\frac{1}{2}+\sigma)}+N^*r^{(\frac{1}{2}-\sigma)}
\label{relation}
\end{eqnarray}
where
\begin{eqnarray}
\nonumber M&=&\frac{\pi}{\sin\pi(1+\sigma)}\frac{1}{\Gamma(\xi^+-\sigma)\Gamma(1+\sigma)}\\ N&=&\frac{\pi}{\sin\pi(1+\sigma)}\frac{1}{\Gamma(\xi^+)\Gamma(1-\sigma)}
\end{eqnarray}
and  $M^*$ and $N^*$ are complex conjugates. Now
\begin{widetext}
\begin{eqnarray}
\lim_{r \to 0} D_\alpha \simeq ( M +e^{i\alpha} M^*)r^{(\frac{1}{2}
+ \sigma)}
+ (N +e^{i\alpha} N^*) r^{(\frac{1}{2}-\sigma)}\,,
\label{da}
\end{eqnarray}
\end{widetext}
Equating the coefficient of  (\ref{psiw1}) with  (\ref{da}) we get
\begin{eqnarray}
\frac{A}{B}=\frac{M +e^{i\alpha}M^*}{N +e^{i\alpha}N^*}\in \mathbb{R}
\end{eqnarray}
This is the energy eigenvalue equation, which is  now function of self-adjoint parameter $\alpha$.  By setting a specific value of  $\alpha$ we can get the energy spectrum of the system. For example, two extremum  solutions  can be analytically found. When the right hand side is zero,
\begin{eqnarray}
E=\omega_B(2n+1-\sqrt{l^2+\lambda^2}-l)\,, n \in \mathbb{N}^+
\end{eqnarray}
and  when the right hand side is infinity
\begin{eqnarray}
E=\omega_B(2n+1+\sqrt{l^2+\lambda^2}-l)\,, n \in \mathbb{N}^+
\end{eqnarray}
Besides the spectrum, the existence of one parameter family of self adjoint extensions has far reaching implications. One possible  implication which is important in the present context is the effect on the non-commutative spatial geometry and fractional  quantum Hall effect.

The non-commutativity of the projected coordinates  $x_P, y_P$ described in (\ref{non4}) will now becomes a family of non-commutative geometries defined by
\begin{eqnarray}
[x_P,y_P]_\alpha= i\Theta(B,\lambda;\alpha)\,,
\label{nonalpha}
\end{eqnarray}
which is our main result in this paper.
It is possible to explicitly evaluate the non-commutativity parameter  $\Theta(B,\lambda;\alpha)$ for our case. To get the projected coordinates one need to construct projection operator. In a specific energy sector $n_0$
the projection operator is
\begin{eqnarray}
\mathcal{O}_P=\sum_{l=0}^{\infty}|n_{0},l\rangle\langle n_{0},l|
\end{eqnarray}
Then the two  projected coordinates on a plane is given by
\begin{eqnarray}
\nonumber x_P &=&\mathcal{O}_Px{\mathcal{O}_P} =\sum_{l,l'=0}^{\infty}\langle n_{0},l'|x|n_{0},l\rangle |n_{0},l'\rangle\langle n_{0},l| \\
y_P &=&\mathcal{O}_Py{\mathcal{O}_P} =\sum_{l,l'=0}^{\infty}\langle n_{0},l'|y|n_{0},l\rangle |n_{0},l'\rangle\langle n_{0},l|
\end{eqnarray}
with
\begin{eqnarray}
\nonumber \langle n_{0},l'|x|n_{0},l\rangle &=&\Omega_{l',l}(\delta_{l',l+1}+\delta_{l',l-1}),\\
\langle n_{0},l'|y|n_{0},l\rangle &=& -i\Omega_{l',l}(\delta_{l',l+1}-\delta_{l',l-1})
\end{eqnarray}
\begin{eqnarray}
\Omega_{l',l}=C_{l'}C_{l}\pi \int dr r {\psi_W}^{*}_{n_{0},l'}{\psi_W}_{n_{0},l}
\end{eqnarray}
The commutator of the relative coordinates then yields
\begin{widetext}
\begin{eqnarray}
[x_P,y_P]_\alpha=\Theta(B,\lambda;\alpha)= 2\sum_{l=0}^{l=\infty}|\Omega_{l,l+1}|^{2}[|n_{0},l+1\rangle
\langle n_{0},l+1|-|n_{0},l\rangle\langle n_{0},l|]\,,
\end{eqnarray}
\end{widetext}
where $\Omega_{l,l+1}$ involves the eigenvalues and therefore in general depend on the self-adjoint extension parameter. The explicit form  can be found as
\begin{widetext}
\begin{eqnarray}
\nonumber \Omega_{l,l+1} = C_{l+1}C_{l}\frac{\pi}{\omega_B}\times \hspace{14cm}\\
\left[\frac{\Gamma(1+\mu+\tilde{\mu}+\varrho)
\Gamma(1-\mu+\tilde{\mu}+\varrho)\Gamma(-2\tilde{\mu})}{\Gamma(\frac{1}{2}-\tilde{\kappa}-\tilde{\mu})
\Gamma(\frac{3}{2}-\kappa+\tilde{\mu}+\varrho)}
{_{3}F_{2}}(1+\mu+\tilde{\mu}+\varrho  ,1-\mu+\tilde{\mu}+\varrho ,\frac{1}{2}-\tilde{\kappa}+\tilde{\mu} ;1+2\tilde{\mu},\frac{3}{2}-\kappa+\tilde{\mu}+\varrho ;1)\right. \nonumber \\
\left.+\frac{\Gamma(1+\mu-\tilde{\mu}+\varrho)\Gamma(1-\mu-\tilde{\mu}+\varrho)\Gamma(2\tilde{\mu})}
{\Gamma(\frac{1}{2}-\tilde{\kappa}+\tilde{\mu})\Gamma(\frac{3}{2}-\kappa-\tilde{\mu}+\varrho)}
{_{3}F_{2}}(1+\mu-\tilde{\mu}+\varrho  ,1-\mu-\tilde{\mu}+\varrho ,\frac{1}{2}-\tilde{\kappa}-\tilde{\mu} ;1-2\tilde{\mu},\frac{3}{2}-\kappa-\tilde{\mu}+\varrho ;1)\right]
\end{eqnarray}
\end{widetext}
where $\kappa=\frac{E+\omega_B l}{2\omega}$,$\mu=\frac{\sqrt{l^{2}+\lambda^2}}{2}$,$\varrho=\frac{1}{2}$,$\tilde{\kappa}=
\frac{E+\omega_B(l+1)}{2\omega_B}$,$\tilde{\mu}=\frac{\sqrt{(l+1)^{2}+\lambda^2}}{2}$.
The explicit form of  $\Theta(B,\lambda;\alpha)$ now depends on the self-adjoint extension parameter and the magnetic field for a fixed interaction strength between the electrons. This two degree of freedom can be exploited to explain fractional quantum  Hall effect.  We use the similar line of arguments as is done for  the composite fermions \cite{jain,jain1}.  We assume that the presence of interaction and therefore  imposition of suitable boundary condition  enforces the system paired electrons to have an integer filling fraction. The filling fraction for the paired system has a filling fraction
\begin{eqnarray}
\nu_\alpha = \frac{N}{M}= \Theta(B,\lambda,\alpha)\frac{N}{A}\,,
\label{fill3}
\end{eqnarray}
As assumed  $\nu_\alpha$ is now some integer, say $\nu_\alpha =p$, which imply that the filling fraction for  the quantum Hall system is
\begin{eqnarray}
\nu = \frac{N}{AB}= \frac{p}{B\Theta(B,\lambda,\alpha)}\,.
\label{fill3}
\end{eqnarray}
Now at a specific magnetic field $B$ the  desired fraction could be obtained by tuning the self-adjoint extension parameter $\alpha$ and thereby making $B\Theta(B,\lambda,\alpha)$ to be the required integer or fraction.


To summarize, we discussed the  problem of system of two electrons on a plane subjected to perpendicular magnetic field which is relevant for the study of quantum Hall effect. We introduced inverse square potential and studied  its effect on the non-commutative structure of the projected coordinates. As a result of one parameter family of self-adjoint extensions of the Hamiltonian we obtained a one parameter family of noncommutative spatial geometry. Using the freedom of the self-adjoint extension parameter we may get the filling fraction of the fraction quantum Hall effects.

{\it Acknowledgement.---}
PRG acknowledges the hospitality of the department of theoretical sciences of Satyendra Nath Bose  National Centre for Basic Sciences, India, where part of the work has been completed. We also thank B. Chakraborty for his comments on the manuscript.
PRG also acknowledges the hospitality of  Physics and Applied Mathematics Unit, Indian Statistical Institute, Kolkata, India, during the visit. DS thanks the Council of Scientific and Industrial Research(C.S.I.R), Government of India, for financial support.

\end{document}